%% file: ComBAO.tex
\documentclass[11pt,a4paper]{article}
\usepackage[left=15mm, top=10mm, right=15mm, bottom=20mm, nohead=true, nofoot=true]{geometry}

\usepackage[utf8]{inputenc}
\usepackage{authblk}
\usepackage{indentfirst}
\usepackage{misccorr}
\usepackage{graphicx}
\usepackage{amsmath}
\usepackage{amssymb}
\usepackage{multicol}
\usepackage{bm}
\usepackage{longtable}
\usepackage{pdflscape}
\usepackage{epstopdf}
\usepackage{multirow}
\usepackage{adjustbox}
\usepackage{amsfonts}
\usepackage{ifthen}
\usepackage{fancyhdr}
\usepackage{setspace}
\usepackage{xcolor}
\usepackage[round,authoryear,comma,sort&compress]{natbib}
\usepackage[toc,title,page]{appendix}
\usepackage[hidelinks]{hyperref}
\usepackage{url}
\usepackage{booktabs}

\hypersetup{
    colorlinks=true,
    linkcolor=green,
    citecolor=blue,
    filecolor=magenta,
    urlcolor=cyan,
    pdftitle={Sharelatex Example},
    pdfpagemode=FullScreen,
}
\fancypagestyle{plain}{\chead{\texttt{\textcolor{brown}{Communications of BAO, Vol. 73, Issue 2, 2026, pp. ???-???}}}}
\title{\textbf{A 15-year K-band view of long-period variables in the central region of M33}}
\author[1, 2]{Hedieh Abdollahi\thanks{hedieh.abdollahi@csfk.org}}
\author[3]{Mina Alizadeh}
\author[2]{Fatemeh Nikzat}
\author[2]{Atefeh Javadi}
\author[3,4]{Yousefali Abedini}
\author[5]{Jacco Th. van Loon}

\affil[1]{\scriptsize Konkoly Observatory, HUN-REN Research Centre for Astronomy and Earth Sciences, MTA Centre of Excellence, Konkoly-Thege Mikl\'os \'ut 15-17, H-1121, Budapest, Hungary}
\affil[2]{\scriptsize School of Astronomy, Institute for Research in Fundamental Sciences (IPM), P.O. Box 1956836613, Tehran, Iran}
\affil[3]{Department of Physics, Faculty of Science, University of Zanjan, 38791-45371, Zanjan, Iran}
\affil[4]{Center for Research in Climate Change and Global Warming (CRCC), IASBS, Zanjan, Iran}
\affil[5]{Lennard-Jones Laboratories, Keele University,
Keele, Staffordshire ST5 5BG, UK}


\begin{document}
\pagestyle{empty}
\newpage
\pagestyle{fancy}
\label{firstpage}
\date{}
\maketitle

\begin{abstract}
Long-period variables (LPVs) on the asymptotic giant branch (AGB) trace the late stages of stellar evolution and the return of mass to the interstellar medium. Their periods range from about 100 to 2000\,d, so a few-year near-infrared campaign covers too few cycles for the slowest pulsators. We combine four independent UKIRT $K$-band data sets obtained between 2003 and 2018 into a homogeneous catalogue of the central $\approx$12\,arcmin$^2$ of M33: 847 stars with 21\,085 measurements over 15.11\,yr. Using 82 calibration stars, we place all data on the 2MASS/Vega system and remove inter-dataset offsets of up to 0.22\,mag. Of the 847 stars, 771 are variable ($\sigma_K>0.10$\,mag), and 385 have peak-to-peak amplitudes $A_K\geq1.0$\,mag, which makes them Mira candidates. Of the 156 stars with Spitzer/IRAC [3.6] photometry, 120 have $\langle K\rangle-[3.6]>0.5$\,mag and are candidate dust-enshrouded AGB stars. The catalogue is the basis for a forthcoming period--luminosity and star-formation history analysis.

\end{abstract}

\emph{\textbf{Keywords:} stars: AGB and post-AGB -- stars: long-period variables --
galaxies: individual: M33 -- infrared: stars -- circumstellar matter --
stellar mass loss}

\section{Introduction}
 
During the AGB phase, low- and intermediate-mass stars can undergo large-amplitude pulsations and lose mass at rates of $10^{-8}$--$10^{-4}$\,M$_\odot$\,yr$^{-1}$ \citep{vanLoon1999,Hofner2018,2020ApJ...894..135S,2022IAUS..366..210J,2025ApJ...991...24B,2026ApJ..1006..131G,mahani2026sedustdustybasedsedfittingpipeline,2023IAUS..373..264M}. As long-period variables (LPVs), they trace both a galaxy's evolved stellar population and the mass it has recently returned to the interstellar medium \citep[e.g.][]{Mahani2025}. Near-infrared observations are particularly useful for identifying these stars in other galaxies. Dusty AGB stars that are faint at optical wavelengths can remain detectable in the K band, where their variability amplitudes are substantially larger than typical photometric errors \citep{Wood1992,Wood1998,Whitelock2003}.

Mira and semi-regular variables have periods of about 100--2000\,d \citep{Wood2000,Ita2004,Goldman2019, Abdollahi2026,2018AJ....156..112Y}. A 3–5 yr campaign may cover only one or two cycles of the longest-period pulsators, making it difficult to resolve period aliases or classify borderline semi-regular variables with confidence \citep{Soszynski2004,Soszynski2009}. Archival data sets can extend the baseline, but zero-point differences between instruments can then look like real variability.
 
M33 is well placed for this kind of study. At 955\,kpc ($\mu=24.9$\,mag; \citealt{Bonanos2006}), its AGB stars can be resolved in the near-infrared, and its inner disc has near-solar metallicity \citep{Rosolowsky2008,Bresolin2011}. UKIRT has observed the centre of M33 since 2003, and these data were used to derive its star-formation history from the LPV population \citep{Javadi2011a,Javadi2011b,Javadi2013,Javadi2015}; the same method has since been applied to other Local Group galaxies \citep[e.g.][]{Abdollahi2023, abdollahi2024, khatamsaz2026studyevolutionngc6822i,2019MNRAS.483.4751H,2011ASPC..445..497J,2021ApJ...910..127N,2021ApJ...923..164S,2024ApJ...972...47A}. Here we summarise the 15-yr $K$-band catalogue of \citet{Alizadeh2026M33}, which combines four of these data sets, and describe the variable stars and dusty AGB candidates it contains.
 
\section{Data and catalogue construction}
 
The first data set is the UIST \citep{RamsayHowat2004} monitoring programme of 2003--2007: 50 exposures over 23 nights, a median of 12 epochs per star and a baseline of 3.80\,yr. It contains 18\,398 sources with a typical precision of 0.02\,mag for $K<17$\,mag \citep{Javadi2011a}. The second is a UFTI \citep{Roche2003} mosaic from August 2005, co-added from 30 frames, for which PSF photometry is presented here for the first time. The third is WFCAM \citep{Casali2007} imaging from 2005--2007, for which we made new PSF photometry of the four central tiles \citep{Alizadeh2025}. The fourth is a single epoch from November 2018 taken from the UKIRT Hemisphere Survey (UHS) DR3 \citep{Schneider2025}.
 
The UIST, UFTI and WFCAM images were reduced with \textsc{daophot/allstar} \citep{Stetson1987,Nikzat2013}; for WFCAM, an empirical PSF was built from isolated bright stars in each frame. The UHS epoch was reduced with \textsc{Photutils} v1.13 \citep{Bradley2024}, part of the Astropy ecosystem \citep{Astropy2022}. A comparison on an archival UFTI H-band frame reduced with both codes gave a mean difference of +0.036\,mag ($\textsc{Photutils}-\textsc{DAOPHOT}$) for 1522 common sources, with a standard deviation of 0.074 mag and a median absolute deviation of 0.048\,mag. All photometry in the catalogue is PSF-based. The central field is very crowded: In the UHS source catalogue, about 62 per cent of detected sources within 1 arcmin of the nucleus belong to blended groups, and aperture photometry in this region is biased by the light of close neighbours \citep{Alizadeh2024,alizadeh2025unveilinglongperiodvariablesm33s,Alizadeh2026M33}. This is why the earlier aperture-based WFCAM catalogue \citep{Javadi2015} is not used.
 
All catalogues were matched to the UIST astrometric frame with a 0.5\,arcsec radius; the median separation is 0.14\,arcsec. We kept stars detected in at least two data sets. The final catalogue has 847 stars and 21\,085 $K$-band measurements over 15.11\,yr (2003 October to 2018 November), and covers $\approx$12\,arcmin$^2$ ($\approx$0.9\,kpc$^2$). Each star has between 4 and 63 epochs, with a median of 21. The main properties are listed in Table~\ref{tab:summary}.
 \begin{table}[t]
\centering
\caption{Summary of the combined $K$-band catalogue \citep{Alizadeh2026M33}.}
\label{tab:summary}
\begin{tabular}{lc}
\toprule
Quantity & Value \\
\midrule
Temporal baseline & 15.11 yr \\
Unique stars & 847 \\
$K$-band measurements & 21\,085 \\
Calibration stars & 82 \\
Largest offset (UHS) & 0.22 mag \\
Variable stars ($\sigma_K>0.10$ mag) & 771 \\
High-amplitude stars ($A_K\geq1.0$ mag) & 385 \\
Stars with Spitzer [3.6] & 156 \\
Dusty candidates ($\langle K\rangle-[3.6]>0.5$ mag) & 120 \\
\bottomrule
\end{tabular}
\end{table}
\section{Photometric homogenisation}

Each data set was reduced independently, so each has its own zero point. Left uncorrected, these offsets would appear as long-term brightness changes. We adopted UFTI as the reference, since it is calibrated to 2MASS/Vega \citep{Skrutskie2006} as in earlier papers of the series. As calibration stars, we selected those detected in all four data sets, with $K<17$\,mag and an rms scatter below 0.20\,mag. This gives 82 stars with $K=13.9$--17.0\,mag.
 
For each of the other data sets, we applied a constant correction equal to the median magnitude difference from UFTI over these 82 stars (uncertainty $\approx$0.015\,mag). We define the offset as \(\Delta K=K_{\rm dataset}-K_{\rm UFTI}\). The measured median offsets are +0.005\,mag for UIST, -0.096\,mag for WFCAM and +0.218\, mag for UHS; we applied corrections of the opposite sign to bring each data set onto the UFTI reference scale. The large UHS value mostly reflects the arbitrary zero point of the \textsc{Photutils} reduction. None of the offsets depends significantly on magnitude.
 
After the correction, the scatter decreased for 58 of the 82 calibration stars (71 per cent). The mean change was $-11.8$ per cent and the median $-15.2$ per cent. The remaining 24 stars are not strictly constant (rms 0.07--0.19\,mag), and for such stars a constant shift can either raise or lower the scatter, depending on which pulsation phases were observed.
 
\section{The variable population}
 \begin{figure*}[t]
\centering
\includegraphics[width=0.95\textwidth]{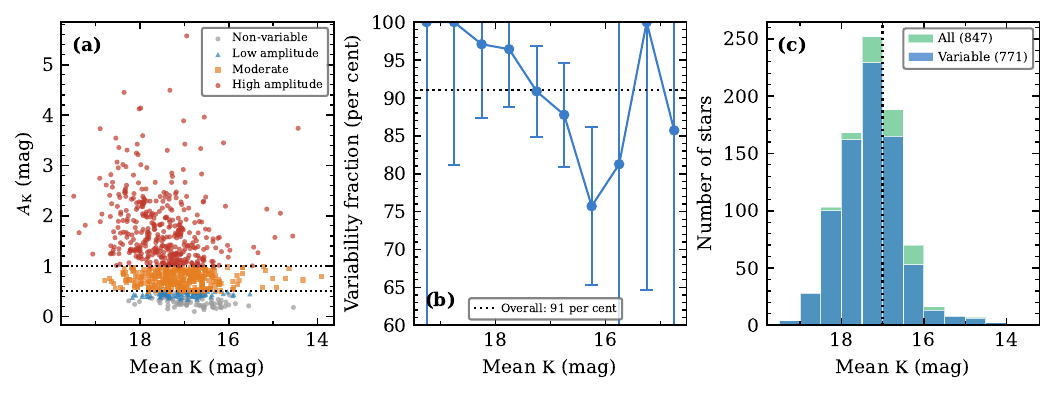}
\caption{Variability statistics as a function of mean $K$ magnitude for the 847 stars. (a) Peak-to-peak amplitude $A_K$, colour-coded by class; the dotted lines mark the class boundaries at 0.5 and 1.0\,mag. (b) Fraction of variable stars in 0.5-mag bins, with $1\sigma$ Poisson errors; the dotted line marks the overall 91 per cent. (c) $K$-band luminosity function of all stars (green) and of the 771 variables (blue); the dotted line marks $K=17$\,mag. Adapted from \citet{Alizadeh2026M33}.}
\label{fig:stats}
\end{figure*}

\begin{figure}[t]
\centering
\begin{minipage}[c]{0.65\linewidth}
  \centering
  \includegraphics[width=\linewidth]{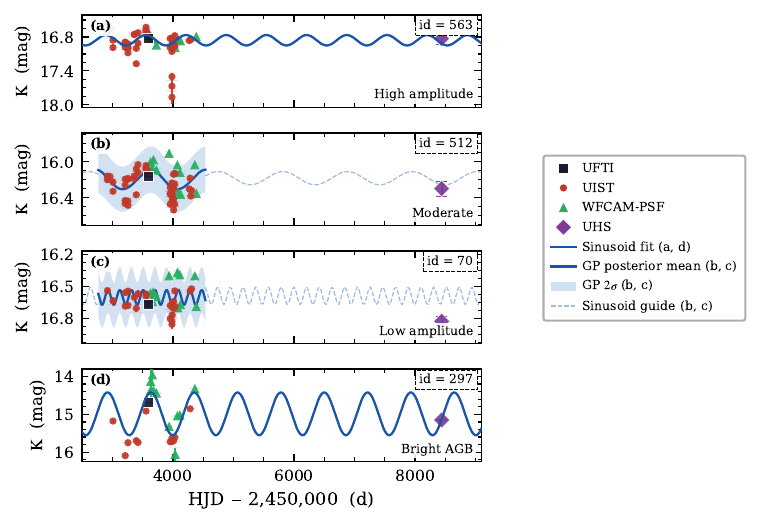}
\end{minipage}\hfill
\begin{minipage}[c]{0.33\linewidth}
\caption{Example $K$-band light curves combining all four data sets ($1\sigma$ errors; the UHS 2018 point is on the right): (a) star 563 ($A_K=2.39$\,mag), (b) star 512 ($A_K=0.63$\,mag), (c) star 70 ($A_K=0.49$\,mag), and (d) the bright AGB star 297 ($A_K=2.13$\,mag, $\langle K\rangle=14.81$\,mag). Curves only guide the eye: a Gaussian-process mean with a $2\sigma$ band in (b) and (c), and a sinusoid at the Lomb--Scargle period in (a) and (d). Figure from \citet{Alizadeh2026M33}.}
  \label{fig:lc}
\end{minipage}
\end{figure}
For each star we computed the scatter of the homogenised magnitudes,
\begin{equation}
\sigma_K =
\left[
\frac{1}{N-1}
\sum_{i=1}^{N}
\left(K_i-\bar{K}\right)^2
\right]^{1/2},
\end{equation}
where $N$ is the number of measurements and $\bar{K}$ is the unweighted mean magnitude. A star is classed as variable if $\sigma_K>0.10$\,mag, about five times the typical precision. This applies to 771 of the 847 stars (91 per cent). The fraction is high because the catalogue contains bright stars detected in several surveys, which are mostly AGB stars; it does not describe the whole stellar population of M33.
 
As a measure of amplitude, we use the peak-to-peak range $A_K=K_{\rm max}-K_{\rm min}$, which does not depend on a period. For the 17 stars with ten or fewer epochs, it is a lower limit. A further 21 stars pass the variability threshold because of a single deviant epoch; they are kept but flagged. We divide the variables into three amplitude classes (Fig.~\ref{fig:stats}a): 385 high-amplitude stars ($A_K\geq1.0$\,mag, mean $A_K=1.77$\,mag), 308 moderate-amplitude stars ($0.5\leq A_K<1.0$\,mag) and 78 low-amplitude stars ($A_K<0.5$\,mag). The other 76 stars are not variable. These classes are descriptive and may change once periods are determined. Example light curves are shown in Fig.~\ref{fig:lc}.
 
The variable fraction lies between 76 and 100 per cent at all magnitudes (Fig.~\ref{fig:stats}b). The luminosity function (Fig.~\ref{fig:stats}c) peaks near $K\approx17$\,mag, or $M_K\approx-7.9$\,mag at $\mu=24.9$\,mag, close to the tip of the thermally pulsing AGB. The catalogue becomes incomplete fainter than $K\approx18$\,mag.
 
\section{Dusty AGB candidates}

 \begin{figure}[t]
\centering
\includegraphics[width=0.8\columnwidth]{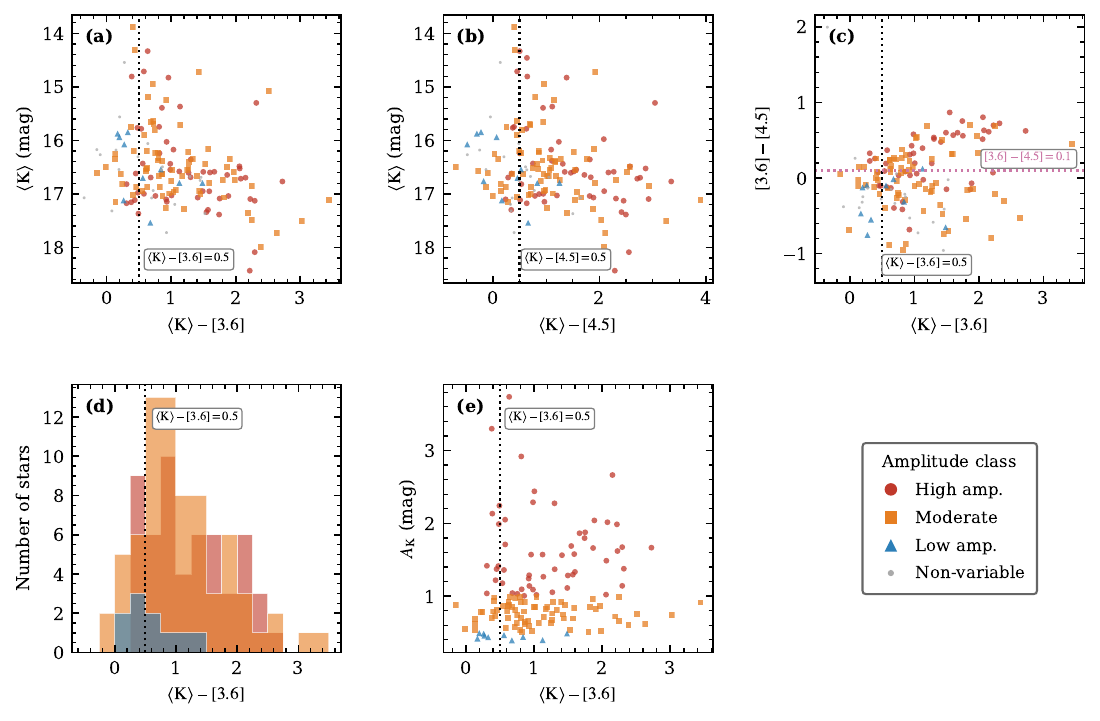}
\caption{Near- to mid-infrared colours of the Spitzer-matched stars, colour-coded by amplitude class: (a) $\langle K\rangle$ vs.\ $\langle K\rangle-[3.6]$; (b) $\langle K\rangle$ vs.\ $\langle K\rangle-[4.5]$; (c) $[3.6]-[4.5]$ vs.\ $\langle K\rangle-[3.6]$; (d) $\langle K\rangle-[3.6]$ distribution per class; (e) $A_K$ vs.\ $\langle K\rangle-[3.6]$. Purple line in (c): DUSTiNGS extreme-AGB limit, $[3.6]-[4.5]=0.1$\,mag \citep{Boyer2015}. Black dotted line: our dusty-candidate threshold, $\langle K\rangle-[3.6]=0.5$\,mag. Figure from \citet{Alizadeh2026M33}.}
\label{fig:dust}
\end{figure}
Circumstellar dust absorbs stellar radiation and re-emits it at infrared wavelengths, making the \(\langle K\rangle-[3.6]\) colour a useful indicator of circumstellar dust emission \citep{SEIP2020,McQuinn2007}. We accepted a match only when the two sources were each other’s nearest neighbours within 1 arcsec. In total, 156 stars have valid [3.6] photometry (114 from SEIP and 42 from \citet{McQuinn2007}).
 
Of these, 120 have $\langle K\rangle-[3.6]>0.5$\,mag and are candidate dust-enshrouded AGB stars (Fig.~\ref{fig:dust}). This number is a lower limit. Only 18 per cent of the catalogue has a Spitzer counterpart. This low fraction is largely attributable to crowding in the nuclear field and should not be interpreted as evidence that the remaining stars lack circumstellar dust. Such stars are hard to reach in the optical: even the brightest star in our sample ($\langle K\rangle=13.88$\,mag) would have $V>18.9$\,mag for $V-K=5$\,mag, below the $V<17$\,mag limit of ASAS-SN \citep{Jayasinghe2018}.

\section{Discussion and conclusions}
 
The main gain is the length of the baseline. For a Mira with $P=500$\,d, UIST alone covers about 2.8 cycles, while the full 15.11-yr series covers about 11. For $P\sim1500$\,d, the combined data cover about 3.7 cycles, whereas UIST alone does not complete one. The 11-year gap between the final WFCAM epoch and the UHS observation extends the baseline and provides a late-time measurement that may help identify long-term changes in stellar brightness.
 
In summary, the four UKIRT data sets, placed on one photometric system, give a 15-yr $K$-band time series for the central region of M33 (Table~\ref{tab:summary}). Most stars in it are variable AGB stars; 385 of them are Mira candidates, and at least 120 are candidate dust-enshrouded AGB stars that can be difficult to identify in optical surveys. Our next step is to determine periods and derive period--luminosity relations, mass-loss rates, and the star-formation history from this sample.


\scriptsize
\bibliographystyle{ComBAO}
\nocite{*}
\bibliography{ComBAO}

\end{document}